\documentclass[reprint,superscriptaddress]{revtex4-2}
\usepackage{graphicx}  
\usepackage{dcolumn}   
\usepackage{bm}        
\usepackage{amssymb}   
\usepackage{amsmath}
\usepackage{xspace}
\usepackage[dvipsnames]{xcolor}
\usepackage{float}
\usepackage [T1]{fontenc}
\usepackage [utf8]{inputenc}
\usepackage[colorlinks]{hyperref}
\usepackage{braket}
\usepackage[paperwidth=210mm,paperheight=297mm,centering,hmargin=2cm,vmargin=2.5cm]{geometry}
\usepackage{multirow}
\usepackage{physics}

\newcommand{\tw}[1]{\ensuremath{t_{w_{#1}}}\xspace}  
\newcommand{\Tg}{\ensuremath{T_{g}}\xspace}                    
\newcommand{\chipp}{\ensuremath{\chi^{\prime\prime}}} 

\begin{document}
\title{Investigating the Interplay Between Aging and Rejuvenation in Spin Glasses}

\author{J. Freedberg} \email{freed114@umn.edu} \affiliation{School of Physics and Astronomy, The University of Minnesota, Minneapolis, Minnesota 55455, USA}

\author{D.~ L.~ Schlagel} \affiliation{Division of Materials
  Science and Engineering, Ames Laboratory, Ames, Iowa 50011, USA}

\author{R.~L.~Orbach}
\affiliation{Texas Materials Institute, The University of Texas at Austin,
  Austin, Texas  78712, USA}

\author{E. Dan Dahlberg}\affiliation{School of Physics and Astronomy, The University of Minnesota, Minneapolis, Minnesota 55455, USA}

\date{\today}

\begin{abstract}
    Aging in a single crystal spin glass ($\mathrm{Cu}_{0.92}\mathrm{Mn}_{0.08}$) has been measured using ac susceptibility techniques over a temperature range of $0.3 - 0.8 \, T_g$.  In these studies, traditional aging experiments (or ``quench'' aging protocols) are compared to aging curves constructed from finite-cooling-rate curves. By comparing the growth rates of spin glass order between the two types of aging curves, it is determined that quantitative comparisons between protocols which are taken by quenching and protocols using a finite cooling rate are not possible without a deeper understanding of the interplay between aging and rejuvenation. We then demonstrate that the data presented indicate that rejuvenation, rather than cumulative aging, is the cause for the discrepancies between the two growth rates.
\end{abstract}

\maketitle

\section{Introduction}
Spin glasses are materials which are inherently time dependent due to their out-of-equilibrium nature. This leads to a wide variety of fascinating phenomena, from the relaxation seen during aging \cite{kisker:96, janus:18, bouchaud:01, kenning:18}, to the reduction in the memory effect \cite{freedberg2023nature, jonsson:04}.  These features also make spin glass intuition widely applicable to other types of disordered systems as exemplified by relaxation or strain-induced memory-like effects in polymers \cite{chen:23, kuersten:17,prados:14, tong:2023strain, jafari:2017,Saleh:2020}.  \\

In metallic spin glasses, such as the $\mathrm{CuMn}$ reported in this paper, the presence of magnetic impurities, in this case the $\mathrm{Mn}$, create frustrated exchange interactions. This is typically attributed to the  RKKY interaction\cite{ruderman_kittel_PhysRev.96.99, yosida_PhysRev.106.893, kasuya1956theory}, where the effective $\mathrm{Mn}$-$\mathrm{Mn}$ interactions may be either ferromagnetic or antiferromagnetic depending on their random locations within the solid solution. While in practice, the frustration mechanism in the actual $\mathrm{CuMn}$ samples tends to be more complicated due to other effects such as $\mathrm{Mn}$ clustering and the possible formation of spin density waves, the glassy behavior expected from random frustrated exchange interactions are still evident in $\mathrm{CuMn}$. In fact, Refs. \cite{sherrington1975solvable,barbara1981scaling,dieny1986critical}, demonstrate multiple disordered interaction pathways which ultimately lead to the glassy behavior expected from only having frustration due to RKKY. Stated simply, the crucial ingredient for spin glass physics within CuMn is the random frustrated exchange interactions between the Mn moments. \\


This frustration, regardless of its mechanism, will lead spin glasses to possess a so-called ``rugged energy landscape.'' In it, there are several nearly-degenerate local minima separated by a large range of barrier heights \cite{vincent_spin_2018,refregier:87, lefloch:92}.  This has a profound effects on the spin glass -- it is unable to explore the entire energy landscape and is thus non-ergodic, which is said to lead to aging \cite{lefloch:92,vincent_spin_2018,refregier:87, bouchaud:01}.  \\

In dc experiments, aging is defined as the process of holding the sample at a constant temperature $T$  below the glass temperature $\Tg$ for some waiting time $t_w$. This allows the spin glass to relax. This can also be seen in ac experiments, where aging appears as a systematic decrease in the susceptibility with respect to time, as shown in the top panel of Fig. \ref{fig:protocol}.  \\

After an aging protocol, if the temperature is subsequently lowered, rejuvenation, defined in an ac setting \cite{jonason:98} can be observed. Rejuvenation is the process where the susceptibility systematically \textit{increases} after an aging protocol, and eventually returns to a ``reference curve'', or a protocol where no aging occurs. Rejuvenation also manifests in dc measurements, and is said to occur when field cooled aging between two temperatures is no longer cumulative. The cause of rejuvenation is typically attributed to the deformation of the energy landscape such that, at some temperature which is lower than \tw{} by $\delta T$, spin glass order must begin to grow anew \cite{bray:78, jonason:00, Jonnason:99,janus:21, miyashita:01, bouchaud:01, jonsson:04}. \\

\begin{figure}
    \centering
    \includegraphics[width = \columnwidth]{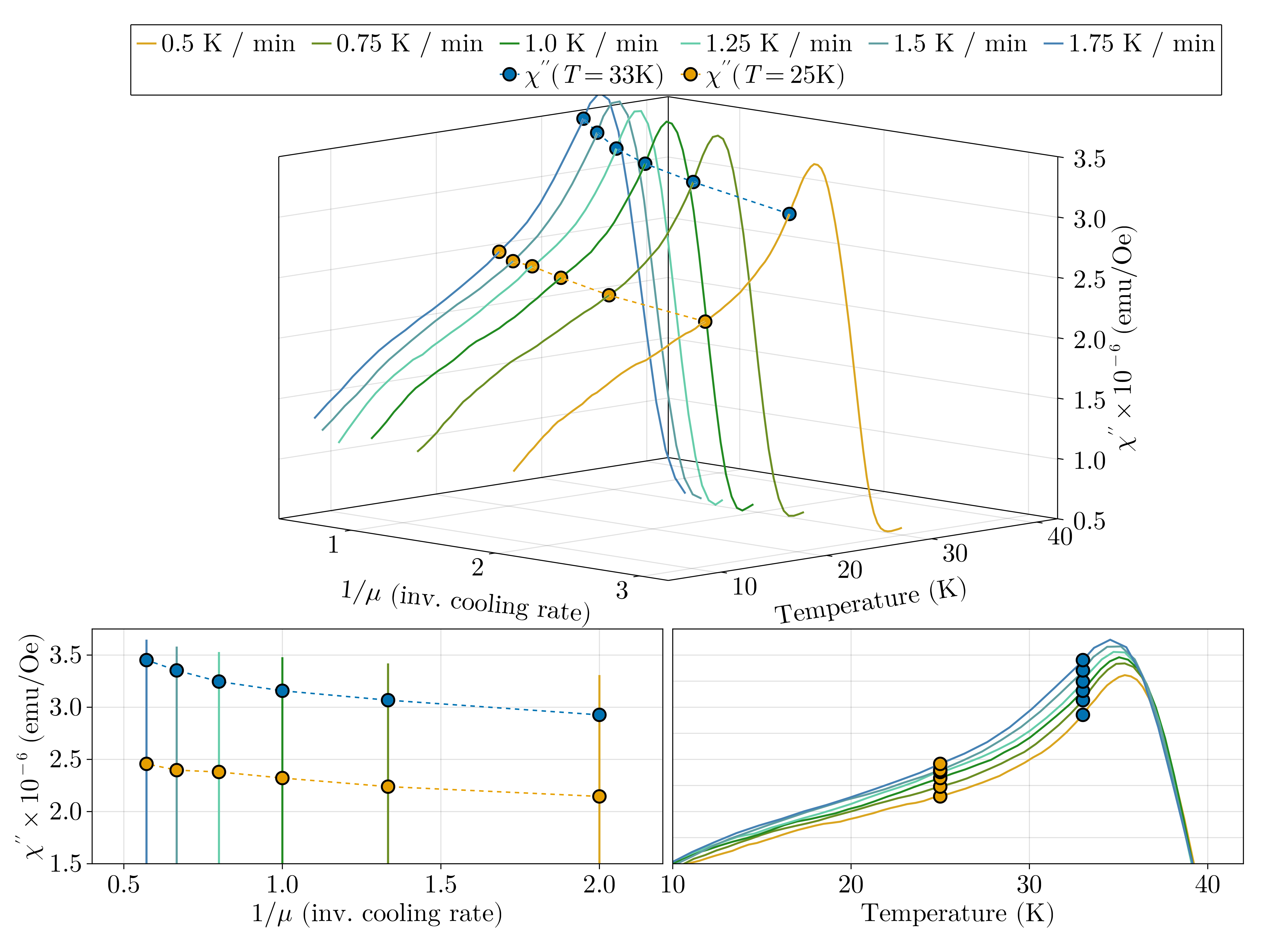}
    \caption{Susceptibility as a function of temperature and inverse cooling rate. Lower cooling rates (and therefore higher inverse cooling rates) are closer to the front of the page. The dashed lines denote constant temperatures to illustrate the decay of the susceptibility between reference curves of different cooling rates. The bottom two panels are cross-sections of each axis.}
    \label{fig:3d_plot}
\end{figure}

However, there is a discrepancy between the rejuvenation seen in dc and ac measurements. In ac measurements, this $\delta T$ is almost an order of magnitude larger than in dc measurements, even within the same sample (or single crystals cut from the same boule). This discrepancy can be seen in Ref. \cite{djurberg:1999} and between Fig. 1 in both Ref. \cite{freedberg2023nature} and \cite{zhai:22}. Thus, there are hints that rejuvenation between dc and ac experiments are not equivalent, and this might be due to the presence of a finite cooling rate in the majority of ac experiments.\\

In most ac susceptibility experiments, a finite cooling rate is used to sweep temperature, and this effect can easily be seen. As shown in Fig. \ref{fig:3d_plot}, the reference curves of \chipp \, versus $T$ with lower cooling rates have systematically lower susceptibilities -- a phenomena which is typically attributed to aging effects \cite{jonason:98}. Traditional wisdom would say that if a sample is not rejuvenating, then it is cumulatively aging. If this is the case, then the susceptibility at equal times in the spin glass state between a reference curve should always be \textit{lower} than the susceptibility from a curve which has been quenched and held at a fixed waiting temperature. However, it is also well-known that rejuvenation occurs as the temperature changes after aging. This means that, as a sample is cooled at a finite rate, \textit{both} aging and rejuvenation must be simultaneously occurring, though the relative proportion of either is unclear. \\

Despite extensive studies characterizing aging, rejuvenation, and memory, the effect that a finite cooling rate has on glassy dynamics is not well-studied. Even ac experiments which investigate different cooling rates ultimately conclude that the effect is essentially negligible over a large temperature range \cite{jonason:98,Jonnason:99}. Thus, finite cooling rate effects, while widely noted across the spin glass literature, are typically neglected.  As such, the interplay between aging and rejuvenation have on glassy dynamics has never been quantitatively explored.\\

In this paper, we quantitatively investigate the effect that a finite cooling rate has on the growth of spin glass correlations. After comparing the growth rates between the two sets of aging curves, we find that finite cooling rate (FCR) aging protocols lead to quantitatively slower magnetic susceptibility relaxation than quench aging protocols.  We then use our results to determine whether the slower growth in the protocols using a finite cooling rate is due to the effects of cumulative aging, or rejuvenation. \\

To accomplish this, we fit quench aging data to a power law, and compare the growth exponent, to FCR aging data. The protocol for constructing both curves is detailed in Fig. \ref{fig:protocol}. A power law fit was chosen as it is the expected behavior of critical fluctuations and has been used to explain dc magnetization experiments and simulations \cite{zhai-janus:20a, paga2021spin}. Additionally, ac experiments using both a hierarchical explanation \cite{freedberg2023nature} and  droplet-like explanations \cite{jonsson:04} find success in attributing the decay in susceptibility to the growth of spin glass order.  The power law fit used is
\begin{align}
    \chipp(t,T) = \alpha - \beta t^{\gamma}, \label{eq:power_law}
\end{align}
where $\chipp(t,T)$ is the measured susceptibility at some temperature and time,  $\chipp_0(0,T) \equiv \alpha$  is the fitted value for the initial susceptibility at $t=0$, and $\gamma$ is the growth rate of the spin glass order. We wish to emphasize that the main point of this Letter is not to determine a functional form for aging, nor is it to test any growth laws. Instead, the aim of this Letter is to test the assumption that all aging curves are the same. We then compare the fitted exponent $\gamma$ from FCR aging curves to quench aging curves. We specifically avoid choosing a form for the correlation length in order to highlight the data -- ultimately, the claim we make in this manuscript is agnostic to \textit{how} spin glass order actually grows. We only assume that growing spin glass order leads to aging.\\

We do, however, use the assumption that rejuvenation is caused by temperature chaos to interpret our results, as described in Refs. \cite{bray:78, zhai:22, miyashita:01, janus:21}. Using a similar argument to what was presented in our previous work, Ref. \cite{freedberg2023nature}, we consider an interplay between the growth of independent correlated regions.  As we cool at some finite rate, the correlation lengths will grow cumulatively, but also go chaotic. The slower the cooling rate, it is expected that there will be more time for cumulative aging to occur, and the spin glass correlation length $\left(\xi(t,T)\right)$ will increase. However, it is unclear how the process of rejuvenation and/or temperature chaos lead to the history-dependent growth of the dynamical correlation length as temperature is varied. Investigating the interplay between these effects will offer valuable insights to the nature of the energy landscape.

\section{Methods}

\begin{figure}
    \centering
    \includegraphics[width = 1\linewidth]{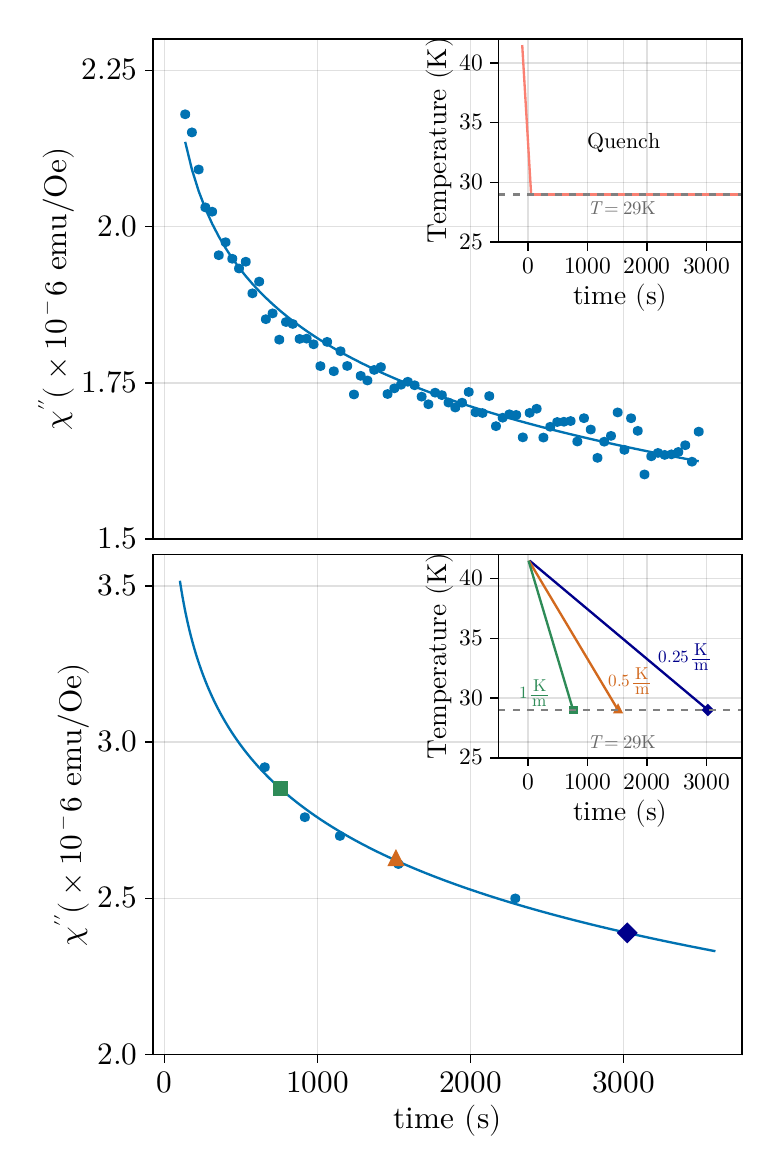}
    \caption{Top: A typical quench protocol and the resulting data. Bottom: An example of constructing an aging curve out of reference curves (FCR aging), as shown in Fig. \ref{fig:3d_plot}. Power law fits of the form Eq. \ref{eq:power_law} are then applied and the growth rates compared. In both the top and bottom panels, the insets show the temperature profile as a function of time.}
    \label{fig:protocol}
\end{figure}

The following measurements were taken on a single crystal of $\mathrm{Cu}_{0.92}\mathrm{Mn}_{0.08}$ using the ac option of a Magnetic Property Measuring System (MPMS) 3 \footnote{Quantum Design North America, 10307 Pacific Center Court San Diego, CA 92121}. The measurement frequency was 1 Hz, with a field amplitude of 10 Oe, which was tested to ensure the spin glass was still in the linear regime. The glass temperature was found through dc magnetization measurements to be $T_g = 41.6$ K \footnote{The value is determined by the ``onset of irreversibility,'' or the temperature where the field cooled and zero-field cooled curves begin to differ from each other. For more details, see Ref. \cite{vincent_spin_2018}}.  \\

An example of how these measurements were taken is shown in Fig. \ref{fig:protocol} where the top panel and inset describes quench aging and the bottom panel and inset describes the process of constructing an FCR curve.  We emphasize that there are only two types of experiments constituting all data shown in this paper. The first set are the quench aging curves, which are compared to FCR curves constructed from reference curves.\\

In the top panel in Fig. \ref{fig:protocol}, for the quench aging, we use a cooling rate of 35 K/min, the fastest cooling rate attainable by the MPMS 3.  For consistency in the measurements, the temperature was allowed to settle at each waiting temperature before recording the first measurement. While the temperature change from $T >T_g$ to the target temperature ranged from only about half to three-quarters of a minute, the first data point taken was around 100 seconds after reaching the target temperature.  Clearly, this means that the true initial value of $\chipp(0,T)$ will not be recorded, so we incorporate this into our fitting procedure as the value $\alpha$. The inset displays the temperature profile as a function of time, and the main plot shows the resulting susceptibility as a function of time. Once the sample reaches the measuring temperature, the temperature is held fixed for the duration of the measurement (in this case, 1 hour). The resulting data are plotted, and the fit applied is the solid line.\\

For the FCR curves, the protocol is depicted by the inset in the bottom panel of Fig. \ref{fig:protocol}, and in Fig. \ref{fig:3d_plot}, which plots the susceptibility as a function of the temperature and the inverse cooling rate. First, the temperature was swept from above $T_g$ to a base temperature below $T_g$ such that the typical temperature range during measurements was $41.5-5$K. This process was repeated for $6$ different cooling rates, ranging from $1.75 - 0.05 \, \mathrm{K/min} $. In the inset, three of the $6$ different cooling rates are highlighted explicitly -- $0.25, 0.5,$ and $1 \mathrm{K/min}$. For each different cooling rate, once the sample reaches the measuring temperature, the susceptibility is recorded. This is represented by the square, triangle, and diamond markers for these labeled rates, respectively. At the measuring temperature, the time in the spin glass state is determined by the cooling rate, and the resulting susceptibility as a function of time is plotted as shown in the main panel. This, too, is fit to the same power law, and the values for the growth rate are then compared.\\

For each recorded point in the quench curves, 10 measurements were averaged. However, due to the fact that the finite cooling rate curves utilize temperature sweeps instead of settling on each measuring temperature, it was imperative that the data was obtained more quickly. Thus, only two data points were averaged to minimize the effect of a changing temperature range. Due to this, the FCR curves are inherently noisier than the data from the quench ages, and a Savitsky-Golay filter was used to smooth the data \cite{savitsky_golay_doi:10.1021/ac60214a047}.\\

There is one final point to add: in aging experiments, typically the \textit{change} in susceptibility is measured. Here, however, we are comparing absolute susceptibilities. So, larger susceptibilities correspond to smaller decays, and smaller susceptibilities correspond to larger decays.

\section{Results and Discussion} 

\begin{figure}
    \centering
    \includegraphics[width=1\linewidth]{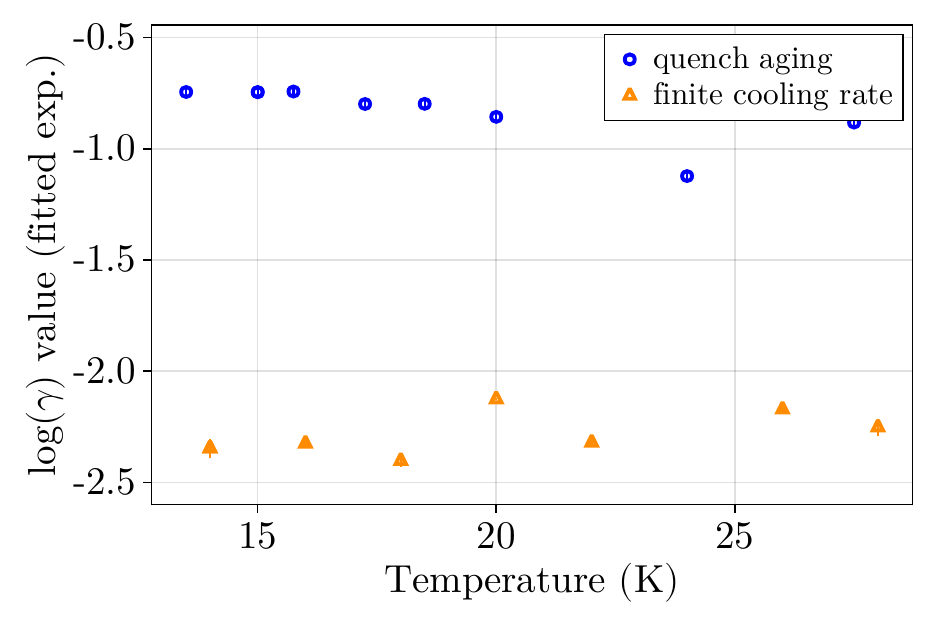}
    \caption{Growth rate $\gamma$ as a function of temperature for quench aging (blue circles) and reference curves (orange triangles). Note that the $y-$axis is a $\log$ scale, and the typical values of $\gamma$ of the curves differ by about two orders of magnitude. 
    \label{fig:gammas}}
\end{figure}

When the procedure outlined in the Methods section and Fig. \ref{fig:protocol} is repeated for many different temperatures using the same power law form of Eq. \ref{eq:power_law}, the fitting parameters can be directly compared to each other. Specifically, we are interested in understanding the differences between the growth rates ($\gamma$s) between quench aging protocols and FCR aging protocols. \\

The results shown in Fig. \ref{fig:gammas} demonstrate that quench aging and FCR aging are quantitatively different from each other. When the growth rate of spin glass order is compared, it is found that not only is this parameter larger for the  quench case than the FCR case, it is orders of magnitude larger. This implies that it is not sufficient to only consider aging effects when describing the results of finite cooling rate protocols. Clearly, something is constraining the correlated regions in FCR curves to grow much more slowly than their quenched counterparts.\\

\begin{figure*}
    \centering
    \includegraphics[width=1\linewidth]{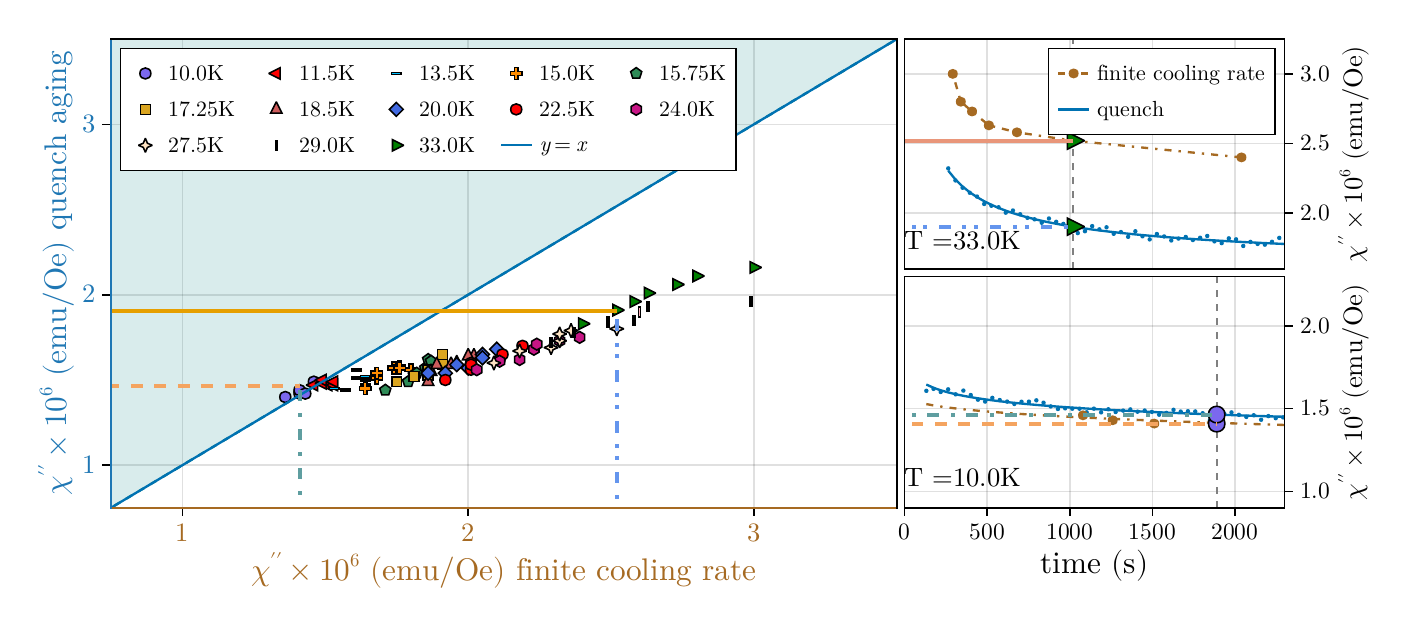}
    \caption{Parametric plot of equal-time susceptibilities for various waiting times. The white region and the teal shaded region are separated by the line $y=x$, which in this case means that the same-time susceptibilities for both the quench aging curves and the finite-cooling-rate curves are the same. However, since almost all of the data points fall below this line, this implies rejuvenation is the dominant cause of the slow growth of spin glass order in finite-cooling-rate experiments.}
    \label{fig:parametric_same_time_susc}
\end{figure*}

The question remains: what is the cause of the slow decay of the FCR curves? There are two possible, but opposite causes. One possibility is that cumulative aging is causing larger (and slower-growing regions) to dominate. Another is that temperature chaos during cooling suppresses the growth of large correlated regions. At the outset, simply studying the growth rates is unable to distinguish between these two cases.\\

To answer this question,  two limiting cases are considered: no chaos (perfect cumulative aging), and perfect chaos (no cumulative aging). We assume the spin glass susceptibility decays faster at a higher temperature than a lower temperature. This is consistent with the temperature dependence of the correlation length growth. Because quench aging experiments are well-characterized in the literature, we will compare the expectations based on each limiting case to the quenches. This will allow us to investigate which effect dominates by determining if the FCR data fall above or below the quench data.\\

First, we consider the case with no temperature chaos and therefore perfect cumulative aging. The susceptibility decay is then completely determined by the accumulative growth of the correlation length at higher temperatures. Thus in this case, at the target temperature, for equal times in the spin glass state, the quenched sample would have a larger susceptibility (and smaller correlation length) than the FCR sample since it was continuously decaying at the relatively higher temperatures. So, if the FCR curves have a \textit{lower} susceptibility than the quench aging curves, cumulative aging is the dominant effect. \\

 In the case of perfect chaos (and therefore no cumulative aging), the FCR curve is continuously ``resetting'' at each infinitesimal temperature step.   In this limit, \textit{any} infinitesimal change in temperature ($\delta T$) will result in the system going chaotic. Thus, between any two temperatures, there is no shared domain structure -- for any measured temperature, the correlation length will start growing anew.   Experimentally, this would mean that there is no decay in the susceptibility as a function of temperature. Once the FCR curve reaches its target temperature, it would have a larger susceptibility than the quench curve  since it has been continuously chaotic and therefore a larger decay rate.\\

To distinguish between the case of only cumulative aging and only temperature chaos, we turn to Fig. \ref{fig:parametric_same_time_susc}, which parametrically compares the susceptibilities at equal times. The two smaller figures on the right show a quench aging curve and a finite cooling rate curve at a high (top) and low (bottom) temperature and one example per plot is provided to demonstrate how the parametric plot is constructed. The grey vertical dashed line indicates the time that the susceptibility is recorded, and their intersection with the blue and orange curves is plotted in the left portion of the figure. For reference, horizontal blue (dot-dashed) and orange (dotted) lines are provided to show where the extracted points are located. The quench aging  susceptibilities are plotted on the $y-$axis is and the FCR susceptibilities are plotted on the $x-$axis. \\

From the plot in Fig. \ref{fig:parametric_same_time_susc}, the two limits can be investigated more easily. Except at very low temperatures where the two are about equal, we see that the reference curve susceptibilities are \textit{always} larger than the quench aging susceptibilities. This can be seen in the top right of Fig. \ref{fig:parametric_same_time_susc}, but more generally, in the main portion of the figure, as almost all of the data points fall below the $y=x$ line. This means that the FCR curves almost always have higher susceptibilities and decay less than the quench curves at equal times.\\

Based on the previous discussion about the limiting behavior, we can see that Fig. \ref{fig:parametric_same_time_susc} rules out the case where cumulative aging dominates. The data indicate that temperature chaos slows the growth of spin glass order during finite cooling protocols.

\section{Conclusions}
In this work, we demonstrate that it is insufficient to just consider aging effects in finite cooling rate protocols. Additionally, we find that the continual ``resetting'' of the spin glass energy landscape as the sample repeatedly rejuvenates is the dominant effect. This causes the apparently slower growth of spin glass order in experiments with finite cooling rates.  These findings clearly demonstrate that when comparing spin glass measurements, the protocol-dependence cannot be ignored. We show that, even if one is cooling at the same average rate, but has a curve which consists of quenches and FCR curves, the two cannot be compared directly. \\

These results can potentially have far-reaching implications for fields outside of spin glasses as well in other glasses which experience rejuvenation after aging, such as in amorphous glasses\cite{utz2000atomistic, dhiraj:10}.  We therefore expect that finite cooling rate effects (or the relevant analog) would be important to consider in any out-of-equilibrium system where temperature changes are continuous.

\begin{acknowledgements}
The authors would like to thank W. Joe Meese, Gregory G.  Kenning, and Samaresh Guchait for many helpful discussions. This work was supported by the U.S. Department of Energy, Office of Basic Energy Sciences, Division of Materials Science and Engineering, under Award No. DE-SC0013599. Crystal growth of $\mathrm{Cu}_{0.92}\mathrm{Mn}_{0.08}$ sample was performed by Deborah L. Schlagel at the Materials Preparation Center, Ames National Laboratory, USDOE and supported by the Department of Energy-Basic Energy Sciences under Contract No. DE-AC02-07CH11358. All experiments were carried out at the Institute for Rock Magnetism (IRM) at the University of Minnesota which is supported by the National Science Foundation, Division of Earth Sciences, Instrumentation and Facilities under Award No. 2153786. 
\end{acknowledgements}

\bibliography{bib}

\appendix
\section{Crystal Growth and Characterization}
Crystal growth and sample preparation was carried out by the Materials Preparation Center (MPC) of the Ames Laboratory, USDOE. Cu from Luvata Special Products (99.99 wt \% with respect to specified elements) and distilled Mn from the MPC (99.93 wt\% with respect to all elements) was arc melted several times under Ar and then drop cast in a water chilled copper mold. The resulting ingot was placed in a Bridgman style alumina crucible and heated under vacuum in a resistance Bridgman furnace to 1050°C, just above the melting point. The chamber was then backfilled to a pressure of 60 psi with high purity argon to minimize the vaporization of the Mn during the growth. The ingot was then further heated to 1300°C and held for one hour to ensure complete melting and time for the heat zone to reach a stable state. The ingot was withdrawn from the heat zone at a rate of 3mm/hr. About 1/3 of the crucible stuck to the alloy. The ingot was finally freed after alternating between hitting with a small punch and hammer and submerging in liquid nitrogen. 

Cross-sections 1-2mm thick were taken from near the start of the crystal growth and from the end for characterization. One side of each was polished and looked at optically and with x-ray fluorescence (XRF). From XRF sample was found to be single phase and the end of the growth to be Mn rich. The samples were then etched in a 25\% by volume solution of nitric acid in water. Optically, the start of the growth is a single phase, single crystal while the end of the growth has large grains with 2\textsuperscript{nd} phase along the grain boundaries. Small pits were seen both optically and by XRF. The pits could be minimized by varying polishing techniques, but not gotten rid of. 

Further investigation was done by polishing the cut ends of the ingot body followed by etching. No evidence of 2\textsuperscript{nd} phase was seen and only occasional small, shallow secondary grains were found. In the Bridgman method, it is not unusual for the very end of the growth to be different because of accumulation of rejected elements and impurities ahead of the growth front. This would account for the change in growth habit (increased number of grains), presence of 2\textsuperscript{nd} phase and overall Mn-rich composition seen at the end of the growth but not in the body. 

\begin{figure}
    \centering
    \includegraphics[width=0.75\linewidth]{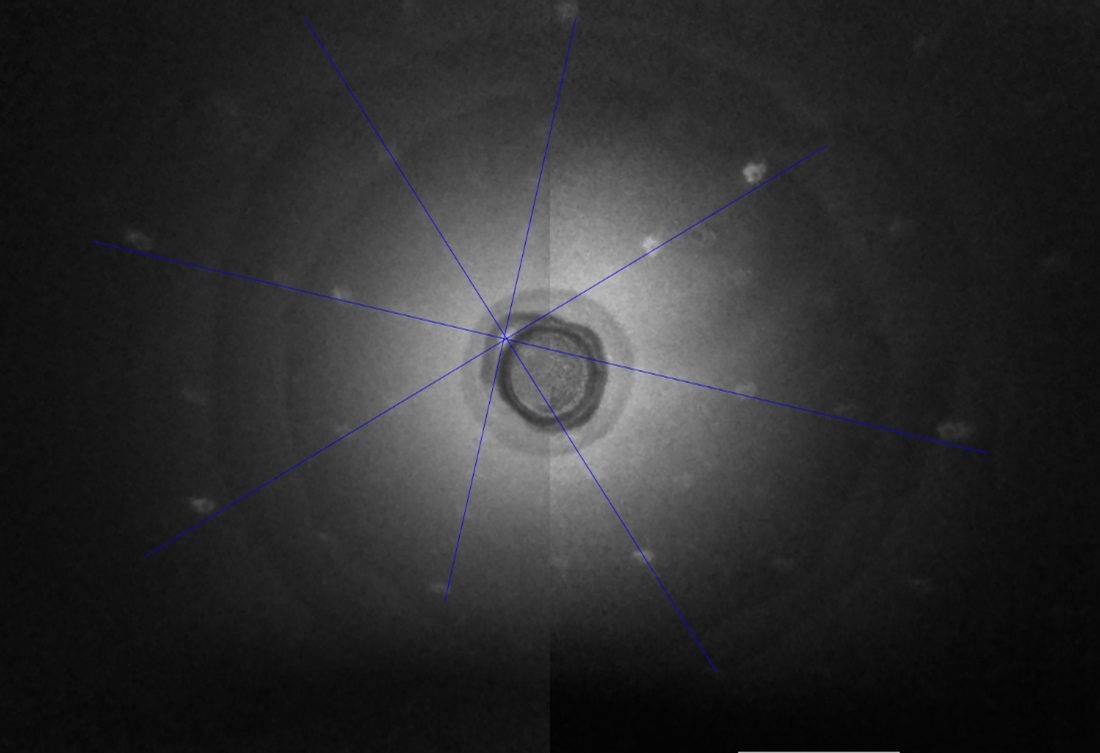}
    \caption{Laue diffraction pattern taken after the crystal was grown.}
    \label{fig:Laue}
\end{figure}

Laue x-ray diffraction along the length of the body confirms that the majority of the body is one single grain. Fig. \ref{fig:Laue}  is a representative Laue image. The spots make a clear pattern which is a qualitative measure of crystal quality. The lines are guides to the eye and reveal that the x-rays are nearly perpendicular to the (100) crystallographic direction. The spot size is proportional to the diameter of the collimator used and each spot appears to make of a cluster of smaller spots. This could be a sign of mosaic spread but since the clusters are fairly tight and overall round in shape, this is not likely a cause of concern. From this image there does not seem to be any evidence of twinning or strain in the crystal. The spots farthest from the center of the image tend to elongate and this is normal. The large concentric shadows are from layers of aluminum foil put on the front of the detector to keep the center of the image from becoming saturated, which are independent of the sample. \\

\end{document}